\documentclass[%
 reprint,
 amsmath,amssymb,
 aps,
]{revtex4-1}
\newcommand{\bi}{\bibitem}
\usepackage{graphicx}
\usepackage{dcolumn}
\usepackage{bm}
\usepackage{braket}
\usepackage{caption}
\usepackage{subcaption}
\usepackage{hyperref}
\usepackage[usenames,dvipsnames]{color}
\newcommand{\cc}{\captionsetup{justification=raggedright,singlelinecheck=false}}

\newcommand{\ct}{\cite}

\begin{document}


\title{Dynamical Quantum Phase Transitions in Extended Transverse Ising Models}
\author{Sourav Bhattacharjee and Amit Dutta \\
Department of Physics, Indian Institute of Technology Kanpur-208016, India}





\date{\today}

\begin{abstract}
We study the dynamical quantum phase transitions (DQPTs) manifested in the subsequent unitary dynamics of an extended Ising model with {additional} three spin interactions{ following a sudden quench}. Revisiting the equilibrium
phase diagram of the model where different  quantum  phases are characterised by different winding numbers, we show {that in some situations the winding number may not change across a gap closing point in the energy spectrum}.  Although, usually there exist{s} a one-to-one correspondence between the change in winding number and the number of critical time scales {associated with DQPTs}, we show
that the extended nature of interactions may lead to unusual situations. Importantly, we show that in the limit
of the cluster Ising model, three critical modes associated with DQPTs  become degenerate and thereby leading
to a single critical time scale for a given sector of Fisher zeros.
\end{abstract}

\maketitle



\section{\label{sec:level1}Introduction}
The study of non-equilibrium dynamics of closed quantum many body systems is an exciting field of recent research both from experimental \ct{greiner02, kinoshita06,fausti11,gring12,trotzky12,cheneau12,rechtsman13,schreiber15} as well
as theoretical viewpoints \ct{calabrese06,rigol08,oka09,mukherjee09,bermudez09,kitagawa10,das10,pal10,lindner11,thakurathi13,Russomanno_PRL12,nag14,patel13,nandkishore15,sen16,bukov16}.  (For review see \cite{dziarmaga10,polkovnikov11,dutta15,eisert15,alessio16,jstat}). One of the exciting features manifested in the subsequent real-time dynamics of   a closed quantum system following a quench,  is the presence of non-analytic behaviour known as the  so-called dynamical quantum phase transitions (DQPTs). The 
notion of DQPTs was introduced by Heyl \textit{et al.},  \cite{heyl13} in  the context of  a one-dimensional transverse Ising chain \cite{sachdev96,suzuki13}; it was observed that 
the dynamical counterpart of the ``free-energy" exhibits  non-analyticities (cusp singularities)   at  some instants of \`critical times' during the subsequent  real time evolution  of the system following a sudden quench  of the transverse field across its quantum critical value.

Let us introduce the basic notion of DQPTs  occurring  in a one-dimensional model following a sudden quench \ct{heyl13};  the system is first prepared in the ground state $|\psi_0\rangle$ of the initial  Hamiltonian $H_i$. One of the parameters of the Hamiltonian is changed suddenly at $t=0$, and the system is then allowed to evolve unitarily  under the new Hamiltonian $H_f$. The Loschmidt overlap amplitude (LOA), defined
as $\mathcal{L}(t)=\bra{\psi_0}e^{-iH_ft}\ket{\psi_0}$, vanishes if the evolved state $|\psi (t)\rangle = e^{-iH_ft}\ket{\psi_0}$, becomes orthogonal to $|\psi_0\rangle$ at some instants of
time, referred to as the critical times; at these instants the so-called dynamical  free energy defined as \ct{gambassi11}

\begin{equation}
f(t)=-\frac{1}{L}\log\mathcal{L}(t),
\label{eq_free_energy}
\end{equation}\\
where $L$ is the linear dimension of the system, exhibits cusp singularities signalling the occurrences of DQPTs. One can  bridge a connection between the DQPTs and classical phase transitions as follows:  An equilibrium classical
phase transition   occur in the thermodynamic limit, when the line of  \textit{zeros} of the generalized  partition function  of the classical system defined on a complex temperature plane (known as Fisher zeros (FZs)) crosses the real temperature axis \ct{lee52,fisher65,saarloos84}.
In a similar spirit, on generalizing the real time $t$ to a complex plane $z={\rm Re} [z] + it$, the LOA takes the  form similar to  the equilibrium (boundary)
partition function $\mathcal{L}(z)=\bra{\psi_0}e^{-H_fz}\ket{\psi_0}$ \cite{fisher65,lee52,saarloos84} and  the corresponding generalized dynamical free energy becomes $f(z)= -(1/ L) \log\mathcal{L}(z) $. The FZs again coalesce into continuous lines in the thermodynamic limit and if these lines cross the imaginary (real time) axis, the non-analyticities in the dynamical free energy, i.e., DQPTs,  are manifested at instants of real time.

 Following the initial proposal, there have been an upsurge of studies probing the possibility of DQPTs in different integrable and non-integrable models \cite{karrasch13,kriel14,andraschko14,canovi14,heyl14,heyl15,budich15,palami15,divakaran16,huang16,puskarov16,zhang16,heyl16,zunkovic16,sei17,fogarty17,heyl18}. (For review see \cite{heyl17,victor17,zvyagin17}). It has also been shown that the occurrence 
of DQPTs is not necessarily entangled with the equilibrium quantum critical point (QCP) for both integrable \ct{vajna14} and non-integrable models \ct{sharma15}. DQPTs have also been observed when the  initial state  is prepared  through a  slow ramping of a parameter of the Hamiltonian from an initial to a final value \cite{pollmann10,sharma16}. Further,  the existence of DQPTs have been established for two-dimensional models, albeit in the form of non-analyticities in the first time derivative of the dynamical free energy  \cite{vajna15,schmitt15,bhattacharya1,bhattacharya2}. Furthermore, an interesting connection between the equilibrium topology and DQPTs have been established in the context of the non-equilibrium dynamics of topological models \ct{vajna15,bhattacharya1}.  Recently, DQPTs have also been shown to exist also when the system is initially prepared in a mixed state \cite{utso,heyl}. Experimentally, DQPTs have been detected in the dynamical evolution of a fermionic many-body state after a quench \cite{flaschner} and also in the nonequilibrium dynamics of a string of ions simulating interacting transverse-field Ising models  \cite{jurcevic16}. 

Unlike conventional order-disorder phase transitions, DQPTs can not be characterised by a local order parameter. However, for an integrable model (reducible to  decoupled two-level systems) the existence of DQPTs can be characterised by a  dynamical topological order parameter (DTOP) derived from  the Pancharatnam phase extracted from the LOA \ct{budich15}. The DTOP sticks to an  integer value as a function of time and exhibits  jumps of unit magnitude at the critical times, thereby characterising DQPTs \cite{budich15}. Finally, an existence of a similar topological order parameter has also been established for slow quenches \ct{sharma16} and also in the case of the two-dimensional Haldane model. 


\begin{figure*}
	\includegraphics[width=\textwidth]{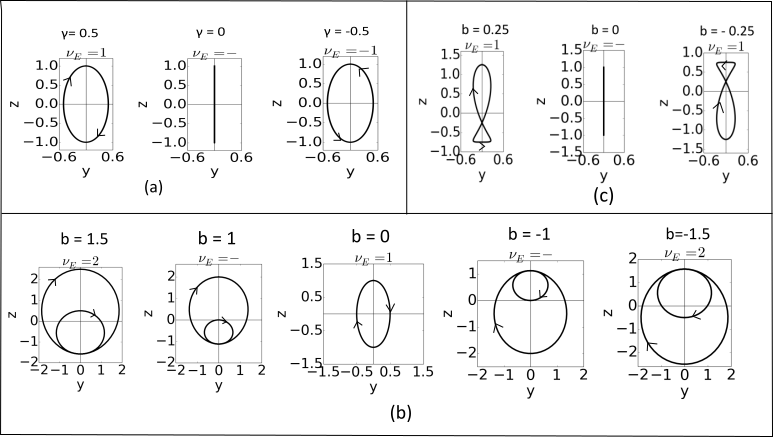}
	\cc
	\caption{Change in $\nu_E$ when (a)  $\gamma$ is varied with $b=0$ (b) $b$ is varied with $\gamma\neq0$. No change in $\nu_E$ (c) when $b$ is varied with $\gamma=0$. The arrows indicate the direction of winding and hence the sign of $\nu_E$. When the loops touch the origin, $\nu_E$ is ill-defined}
	\label{fig_eq_phase}
\end{figure*}

In this paper,  we study DQPTs following a sudden quench of the Hamiltonian for an extended transverse Ising (ETI) chain \cite{zhang,zhang17} which incorporates an additional three-spin interaction term in such a way that the spin-chain is still integrable in terms of Jordan-Wigner fermions.  The generalized versions of the quantum XY/Ising Hamiltonians have been studied extensively both in the equilibrium \cite{titvinidze03, fabrizio96, suzuki71, perk75,divakaran13} and non equilibrium \cite{chowdhury10,divakaran16} situations. Remarkably, Zhang \textit{et al.}, \cite{zhang} showed that the ground states of the Hamiltonian in this model can be represented through loops in a two-dimensional auxiliary space spanned by the parameters of the Hamiltonian. As for example, the loops in the case of the one-dimensional nearest neighbour  transverse Ising (TI) model and XY model take  the shape of a circle and an ellipse in this auxiliary space, respectively. Further, Ref. \cite{zhang} introduced the notion of a \textit{topological} winding number $\nu_E$, defined as the number of times  a particular loop winds around the origin. Evidently,   $\nu_E$  assumes  only integer values  characterizing the equilibrium quantum phases and becomes ill-defined at the QCPs separating two phases. For the nearest neighbour Ising or XY model, the magnitude of the winding number is restricted between the values 
$\pm 1$ while the extended nature of interaction in the ETI chain leads to  quantum phases with integer winding number exceeding $\pm 1$. 
This rich phase diagram enables us to look for DQPTs for quenches across QCPs between two phases whose $\nu_E$ differs by a magnitude as large as $4$. The situation
here resembles the long-range Kitaev chain  \ct{vodola14,vodola16, viyuela16, regemortel16, lepori17, dutta17} where the long-range nature of interaction  lead to topological phases with higher winding numbers in the equilibrium situation. \ct{jaeger17, sedlmayr17}. We note that long-range interacting spin chains \ct{dutta01} has already been addressed in several works \ct{halimeh17, homri17, zunkovicb}.


We summarize the main results of our paper at the outset.  Concerning the equilibrium phase diagram of the model,  we demonstrate a specific case in which the winding number  remains invariant upon changing a particular parameter even though the gap in the spectrum closes at a specific parameter value; however, quenching across the gapless point results in DQPTs. We then establish  the presence of DQPTs for quenches across QCPs where $\nu_E$ changes by $\Delta\nu_E=1,2,3,4$.  We show numerically that the number of critical time scales (at which the free energy becomes non-analytic)  is equal to $\Delta\nu_E$.
Remarkably, we also establish special situations where this ``thumb rule" fails; similar situation have been observed previously in the XY model \cite{vajna14,vajna15}. More importantly,
focussing on the limit in which the ETI reduces to a {\it cluster Ising chain} \cite{son11,smacchia11}, we show  there exist three critical modes within the effective Brillouin zone following a quench across the QCP; remarkably, unlike previous studies \ct{dutta01,jaeger17} these modes are degenerate in the final energy spectrum and hence one obtains a single critical time scale. This degeneracy, which is not accidental, has not been reported in earlier studies to the best of our knowledge.   

The rest of the paper is organised as follows. In section \ref{sec:ETIM}, we introduce the Hamiltonian for the ETI model focussing on some specific cases to demonstrate the possible invariance of winding number across a gapless point. In section \ref{sec:DQPT}, we numerically study DQPTs and associated critical time scales for quenches across QCPs with $\Delta\nu_E=1,2,3,4$. The emergence of degenerate critical modes in the cluster Ising limit of the ETI model is discussed in section \ref{sec: DCM}. Finally, we summarize our results in section \ref{sec:con}.


\section{\label{sec:ETIM}Extended Transverse Ising Model}
The Hamiltonian for a one-dimensional extended transverse Ising model is given as \cite{zhang}:
\begin{multline}
H=\sum_{i=1}^N\left[a\left(\frac{1+\gamma}{2}\sigma_i^x\sigma_{i+1}^x+\frac{1-\gamma}{2}\sigma_i^y\sigma_{i+1}^y\right)+g\sigma_i^z\right]\\+\sum_{i=1}^N\left[b\sigma_i^z\left(\frac{1+\delta}{2}\sigma_{i-1}^x\sigma_{i+1}^x+\frac{1-\delta}{2}\sigma_{i-1}^y\sigma_{i+1}^y\right)\right],
\label{eq_hamil}
\end{multline}
where $\sigma_i^{\alpha}$s, with $\alpha=x,y,z$'s are  Pauli spin operators defined on the lattice site $i$. Here, the parameter $g$ denotes the transverse field while $a$ and $b$ stand for the strength of the nearest
neighbour and the three-spin interaction; these interactions are anisotropic when the anisotropy parameters $\gamma$ and $\delta$, in the nearest neighbour and the three-spin interaction, respectively, are non-zero. 
For $b=0$, the model \eqref{eq_hamil} reduces to a transverse XY chain for any $\gamma \neq \pm1$ and to a TI chain for $\gamma=\pm1$.

Assuming periodic boundary conditions, this Hamiltonian can be exactly solved using the  Jordan-Wigner transformation  that maps the spin Hamiltonian to  a system of spinless fermions. Since the parity of the number of fermions is conserved, the Hamiltonian is block-diagonal with two sectors $H^+$ and $H^-$ for even and odd number of fermions respectively.
\begin{equation}
H=\begin{pmatrix}
	H^+ & 0\\
	0 & H^-
	\end{pmatrix}
\label{eq_hamil_block}
\end{equation} 
Finally, taking the Fourier transform of the fermionic operators in $H^+$ sector,  allows one to decompose $H^+$ into a sum of decoupled two-level Hamiltonians $H(k)$ 
\begin{equation}
H^+=\sum_{k>0}H_k=4\sum_{k>0}\vec{r}(k).\vec{s}(k)
\label{eq_hamil_decouple}
\end{equation}
 
where  $\vec{r}(k)=(x(k), y(k), z(k))$ is given as:

\begin{equation}
\begin{split}
&x(k)=0\\
&y(k)=a\gamma\sin k+b\delta\sin 2k\\
&z(k)=a\cos k + b\cos 2k -g
\label{eq_rk}
\end{split}
\end{equation}
while $\vec{s}(k)$ ($s_x(k),s_y(k),s_z(k)$) represent a set of pseudospins, given in terms of the fermionic creation and annihilation operators \ct{zhang}, corresponding to  momentum $k$. From Eq.~\eqref{eq_rk}, it is obvious that $\vec{r}(k)$ traces out a loop in the  $(y,z)$ plane as the momentum $k$ sweeps over the brillouin zone.
The ground state energy is obtained as:
\begin{equation}
\epsilon_g=-\frac{1}{2\pi}\int_{-\pi}^{\pi}|\vec{r}(k)|dk
\label{eq_energy}
\end{equation}


and its first derivative can be expressed as:
\begin{equation}
\partial\epsilon_g=-\frac{1}{2\pi}\int_{-\pi}^{\pi}\hat{r}(k).\partial\vec{r}(k)dk
\label{eq_energy_deriv}
\end{equation}
Since the unit vector$(\hat{r}(k))$ is ill defined at the origin, the first derivative of the ground state energy becomes non-analytic and the system undergoes a quantum phase transition at $|\vec{r(k)}|=0$. Also, since $\epsilon_g(k)=-|\vec{r}(k)|$, this QCP is naturally associated with the closing of the energy gap in the spectrum for the momentum $k$.  Hence, a winding number can be defined on the auxiliary space i.e., the $(y,z)$ plane,
\begin{equation}
\nu_E=\frac{1}{2\pi}\oint_c\frac{1}{r^2}(ydx-xdy)
\label{eq_winding_number}
\end{equation} 
where $c$ represents the closed loop traced out by $\vec{r}(k)$. This winding number takes on integer values except at the QCP ($r=0$) where it is ill defined. Thus it can serve as a `\textit{topological}' order parameter (TOP) whose value characterizes a quantum phase;  a change in $\nu_E$ usually signals a QPT.  

However, we would like to point out that QPT or rather a closing of gap in the spectrum is not necessarily associated with a change of the $\nu_E$ defined in Eq.~\eqref{eq_winding_number}. To illustrate these points, we present the following three cases (Fig.~\ref{fig_eq_phase}) choosing $a=\delta=1$ and $g=0$ in the Hamiltonian \eqref{eq_hamil}. In the first two cases, the $\nu_E$ changes whenever we cross a QCP. In the third case however, we show that $\nu_E$ shows no change after crossing a QCP, though it becomes ill-defined at the QCP. In all cases, the $\nu_E$ is determined from Eq.~\eqref{eq_winding_number} or equivalently by counting the number of times the loops in the $(y,z)$ plane wind around the origin in the clockwise sense.
\begin{figure*}
	\begin{subfigure}{0.5\textwidth}
		\includegraphics[width=\columnwidth]{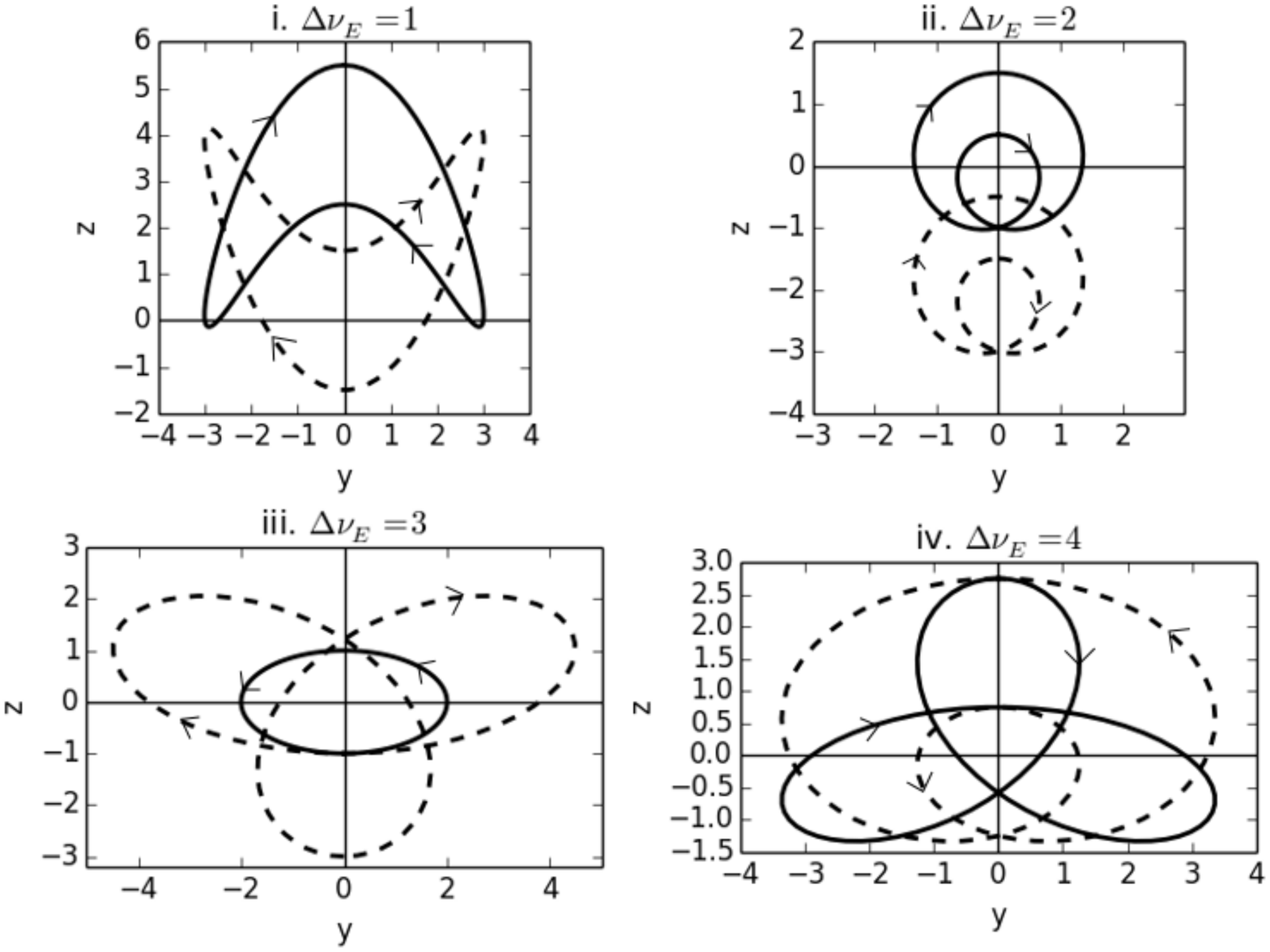}
		\caption{}
		\label{fig_loops}
	\end{subfigure}\quad\quad
	\begin{subfigure}{0.45\textwidth}
		\includegraphics[width=\textwidth]{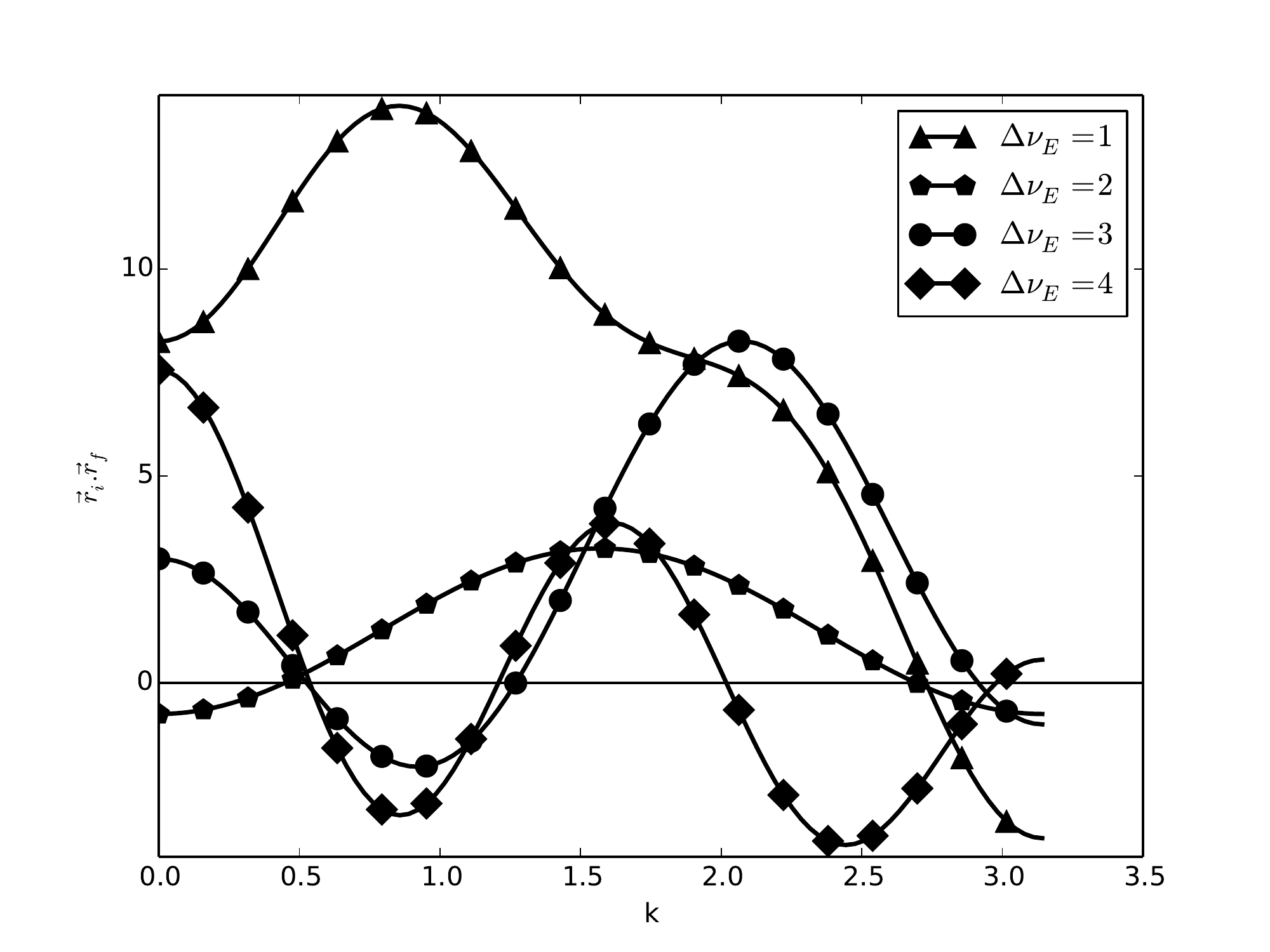}
		\caption{}
		\label{fig_fisher}
	\end{subfigure}
	\cc
	\caption{(a) Loops in auxiliary space corresponding to the ground state of initial (dashed line) and final (solid line) Hamiltonian for the quenches outlined in Table 1. (b) $\vec{r}_i(k)\cdot\vec{r}_f(k)$ as a function of $k$ for each of the quenches. Here, $\vec{r}_i(\vec{r}_f)$ refers to the parameters of the initial(final) Hamiltonian and $\vec{r}_i(k^*)\cdot\vec{r}_f(k^*)=0$ for the critical modes $k^*$. The number of $k^*$ (and hence critical time scales $t^*$) bear a one-to-one correspondence with the corresponding change in $\nu_E$ }
\end{figure*}

\noindent Case a: $b=0$, $\gamma$ is varied: For this case,  the Hamiltonian reduces to the XY Hamiltonian. As $\gamma$ changes from positive to negative values, the corresponding $\nu_E$ changes from $-1$ to $+1$ as shown in Fig.~\ref{fig_eq_phase}a. The anisotropy transition at $\gamma=0$ in the XY model is thus clearly detected by a change of $\Delta\nu_E=2$.\\

\noindent Case b: $\gamma\neq0$, $b$ is varied: There are two critical points at $b=\pm1$. For $b>1$ and $b<-1$, $\nu_E=2$ while $\nu_E=1$ for $-1<b<1$ (see Fig.~\ref{fig_eq_phase}b). Thus $\nu_E$ changes from $2$ to $1$ and again from $1$ to $2$ as the two critical points are crossed by varying $b$.\\%

\noindent Case c: $\gamma=0$, $b$ is varied:  As can be seen from Fig~\ref{fig_eq_phase}c, $\nu_E=1$ for both $b<0$ and $b>0$, even though the band gap closes at the QCP ($b=0$)  where $\nu_E$ becomes ill-defined.  We note in passing that the point ($\gamma=b=0$) represents a multicritical point in the phase diagram of the Hamiltonian \eqref{eq_hamil}.

\section{\label{sec:DQPT}Non Equilibrium Quantum Phase Transition}

To study the non-equilibrium dynamics of the ETI chain, we initially prepare the system in the ground state $\ket{\psi_i^0}$ of the initial Hamiltonian $H_i(k)$ for the mode $k$. The LOA now can be written in the form \ct{heyl13}
\begin{equation}
\mathcal{L}(z)=\prod_{k>0}\mathcal{L}_k(z)=\prod_{k>0}
\bra{\psi_i^0(k)}e^{-H_f(k)z}\ket{\psi_i^0(k)},    \label{eq_LOA_decom}
\end{equation}
where $H_f(k)$ is the final Hamiltonian for the mode $k$. Rewriting Eq.~\eqref{eq_rk} as $(y_{i(f)}(k),z_{i(f)}(k))=(|\vec{r}_{i(f)}(k)|\sin\theta_{i(f)}(k),|\vec{r}_{i(f)}(k)|\cos\theta_{i(f)}(k))$  and expanding $\ket{\psi_i^0(k)}$ in the eigen basis of $H_f(k)$, we obtain
\begin{equation}
\mathcal{L}(z)=\prod_{k>0}\left(\cos^2(\phi_k)e^{\epsilon_f(k)z}+\sin^2(\phi_k)e^{-\epsilon_f(k)z}\right),
\label{eq_LOA_expr}
\end{equation}
where $\epsilon_f(k)$ is the energy of the excited state corresponding to $H_f(k)$ and 
\begin{equation}
\phi_k=\frac{\theta_f(k)-\theta_i(k)}{2}
\label{eq_phi}
\end{equation}

\begin{figure*}
	\includegraphics[width=0.8\textwidth]{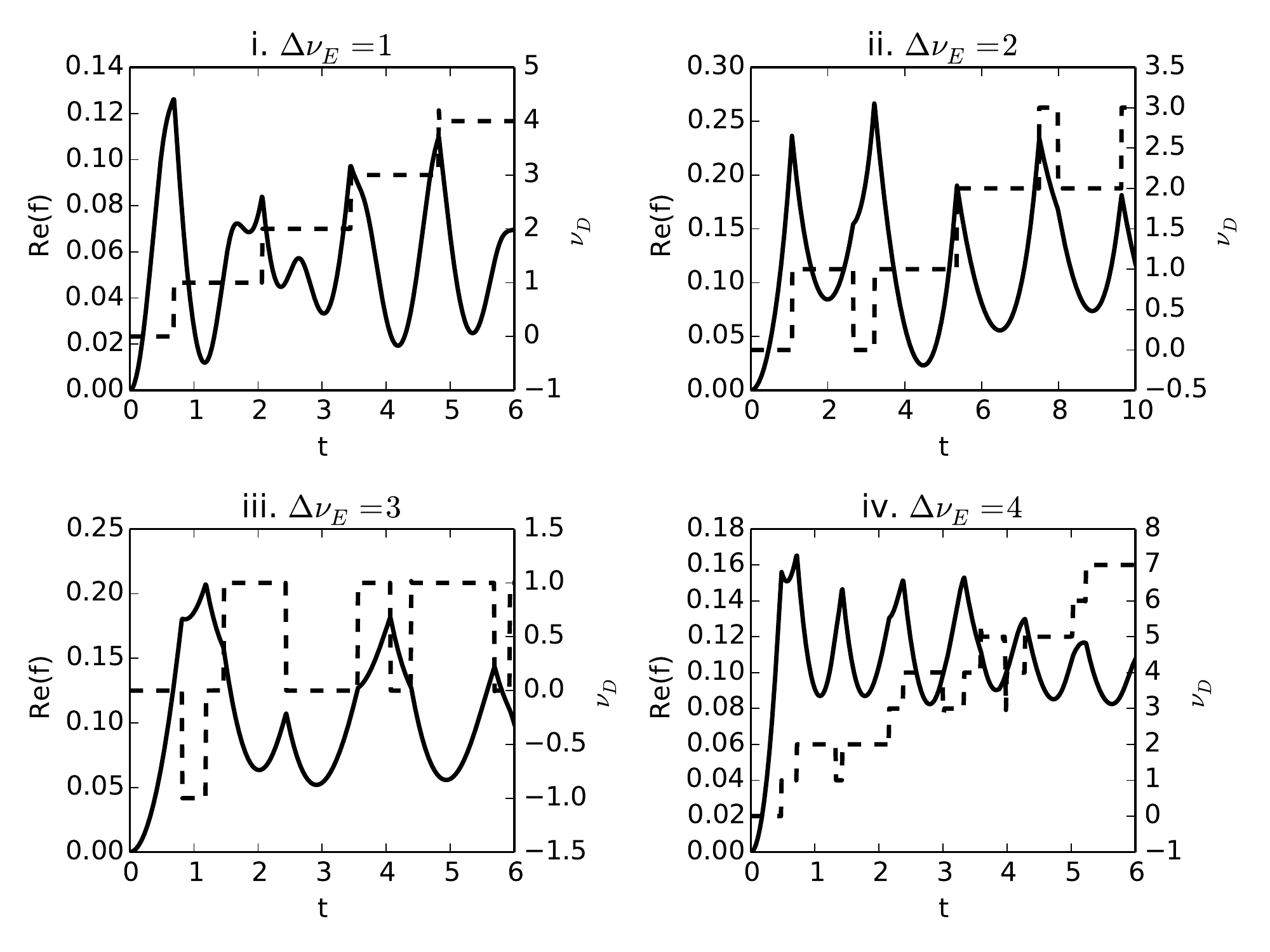}
	\cc
	\caption{Real time evolution of the real part of the dynamical free energy (solid line) and the DTOP (dashed line) following the quenches illustrated in Fig.~\ref{fig_loops}.} 
	\label{fig_dtops}
\end{figure*}
Substituting the expression for the LOA derived above into Eq.~\eqref{eq_free_energy}, we get the dynamical free energy in the thermodynamic limit as 
\begin{equation}
f(z)=-\int_0^\pi\frac{dk}{2\pi}\log\left(\cos^2(\phi_k)e^{\epsilon_f(k)z}+\sin^2(\phi_k)e^{-\epsilon_f(k)z}\right)
\label{eq_free_energy_expr}
\end{equation}

\noindent Evidently, the  Fisher zeros i.e., the zeros of the argument of the logarithm in Eq.~\eqref{eq_free_energy_expr} residing on the complex  $z$ plane, are given by the equation:

\begin{equation}
z_n(k)=\frac{1}{2\epsilon_f(k)}\left(\log{\tan^2{\phi_k}}+i\pi(2n+1)\right)
\label{eq_fisher}
\end{equation}
where $n=0,1,2,...$. For certain critical momenta $k^*$, the real part of $z_n(k^*)$ vanishes which is equivalent to the condition
\begin{equation}
\vec{r}_i(k^*)\cdot\vec{r}_f(k^*)=0,
\label{eq_critical_k}
\end{equation}
The Fisher lines therefore, cross the imaginary (real time) axis at critical times given by the imaginary part of $z_n(k)$:
\begin{equation}
t_n=t^*(2n+1)
\label{eq_critical_t}
\end{equation}
where 
\begin{equation}
t^*=\frac{\pi}{2\epsilon_f(k^*)} 
\label{eq_cr_tscale}
\end{equation}

is the critical time scale when the first DQPT occurs.

Although a DQPT cannot be characterized by a local order parameter, Budich \textit{et al.,} \cite{budich15}  showed that at least for integrable decoupled two level systems, DQPTs can be characterized by a dynamical topological order parameter (DTOP). This DTOP is derived through the gauge invariant Pancharatnam geometric phase (PGP), which is extracted from the phase of the LOA rendered invariant by subtracting the dynamical phase from the same. This PGP changes value by $\mod 2\pi$ as the crystal momentum $k$ sweeps across the brillouin zone. At the critical momenta $k^*$ however, this PGP remains pinned to a value of $0$ or $\pi$ at all times except at $t_n$ for which it is ill-defined. Using the above information, one can define the DTOP as
\begin{equation}
\nu_D(t)=\frac{1}{2\pi}\oint\frac{\partial\phi_k^G(t)}{\partial k}dk,
\label{eq_dtop}
\end{equation}
which remains fixed at constant integer values in between the critical times, thus characterizing the dynamical phase transitions.

\begin{figure*}
	\begin{subfigure}{0.3\textwidth}
		\includegraphics[width=0.8\columnwidth]{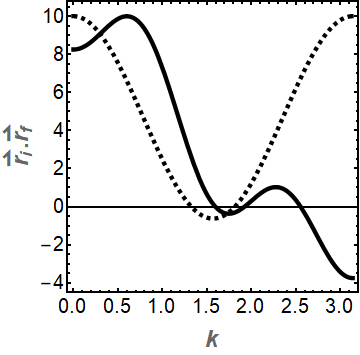}
		\caption{}
		\label{fig_except1}
	\end{subfigure}
	\begin{subfigure}{0.3\textwidth}
		\includegraphics[width=\columnwidth]{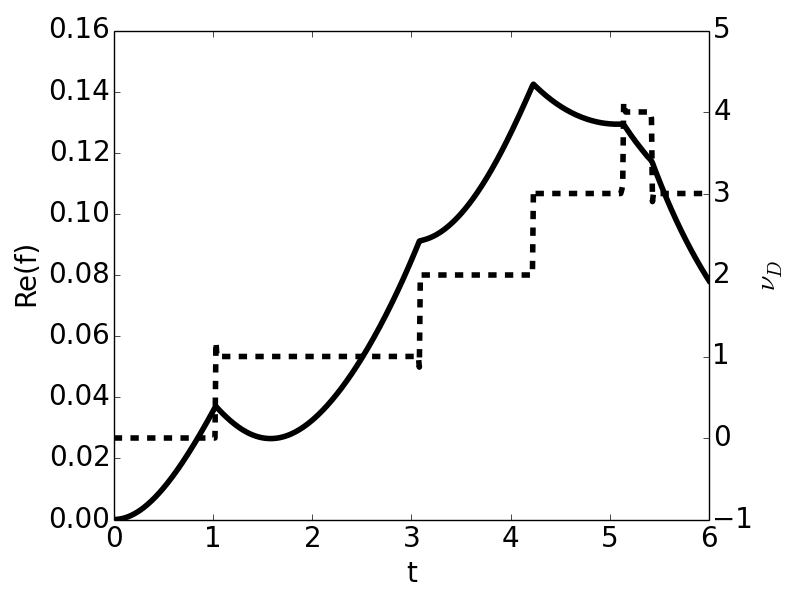}
		\caption{}
		\label{fig_except2}
	\end{subfigure}
	\begin{subfigure}{0.3\textwidth}
		\includegraphics[width=\columnwidth]{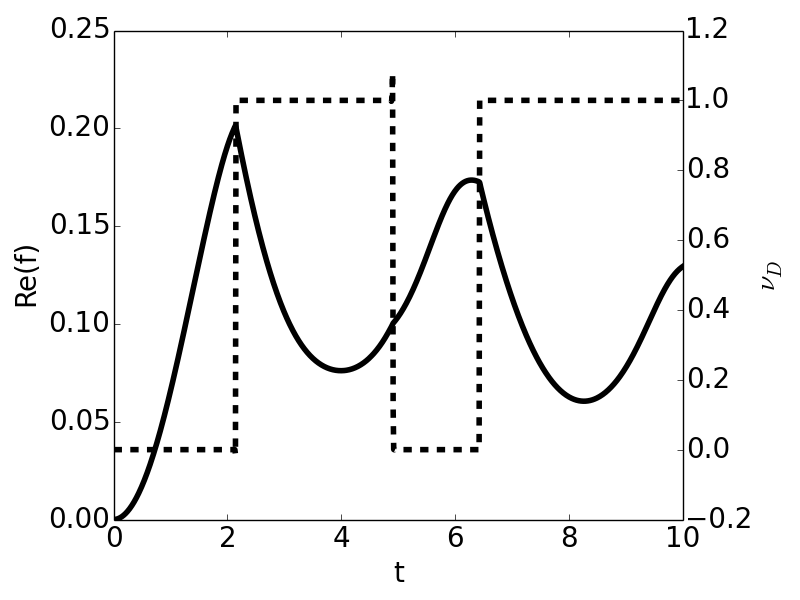}
		\caption{}
		\label{fig_except3}
	\end{subfigure}
	\cc
	\caption{(a) $\vec{r}_i(k)\cdot\vec{r}_f(k)$ plot shows two (dotted line; scaled by a factor of $10$) and three (solid line) $k^*$ even though $\Delta\nu_E=1$ and $=0$ respectively. Real time evolution of free energy and DTOP for (b) showing
		three critical time scales even when $\Delta\nu_E=1$ and (c) showing existence of DQPTs with $\Delta\nu_E=0$}. 
\end{figure*}

We now analyze DQPTs for sudden quenches between the equilibrium phases with different $\nu_E$ as shown in Fig.~\ref{fig_loops}. 
and tabulated below. In all cases, one of the parameters of the Hamiltonian \eqref{eq_hamil} is changed retaining others at fixed values ensuring a change in $\nu_E$.
\begin{center}
	\captionof{table}{}
\begin{tabular}{|c|c|c|c|c|c|c|}
\hline
Case & $a$ & $\gamma$ & $b$ & $\delta$ & $g$ & $\Delta\nu_E$\\
\hline
i & $1.5$ & $2$ & $-2\rightarrow 2$ & $0.05$ & $-2$ & $1$ \\
ii & $-0.5$ & $1$ & $1$ & $1$ & $2\rightarrow 0$ & $2$\\
iii & $-1$ & $-2$ & $-2\rightarrow 0$ & $1.5$ & $0$ & $3$\\
iv & $1$ & $-1.5$ & $1.5$ & $-1.5\rightarrow 1.5$ & $-0.25$ & $4$\\
\hline
\end{tabular}
\end{center}
In Fig.~\ref{fig_fisher}, we plot $\vec{r}_i(k)\cdot\vec{r}_f(k)$ for the above quenches as a function of $k$. As dictated by Eq.~\eqref{eq_critical_k}, the zeros of this function for $0\leq k\leq\pi$ determines the critical  $k^*$. Inspecting Fig.~\ref{fig_loops} and Fig.~\ref{fig_fisher},  we find that the number of $k^*$ are determined by corresponding change in $\nu_E$. Interestingly, as we shall demonstrate later, this is not always the case. The real time evolution of the dynamical free energy and the DTOP for the  four quenches mentioned above is illustrated in Fig.~\ref{fig_dtops}. We analyze them as under \\
\noindent i. $\Delta\nu_E=1:$ There is only one $k^*$, as is evident from Fig.~\ref{fig_fisher}, and the critical times are determined by Eq.~\eqref{eq_critical_t}. The cusps in the free energy and integer jumps in DTOP at these critical times are visible in Fig.~\ref{fig_dtops}i.\\ 
\noindent ii. $\Delta\nu_E=2:$ In this case, Fig.~\ref{fig_fisher} suggests two $k^*$ and hence two sets of critical times. Fig.~\ref{fig_dtops}.ii. shows that while the DTOP jumps to a higher value at the first set of critical times, it lowers for the other set. Whether the DTOP increases or decreases from its initial value at a later time is thus determined by the shortest $t^*$ or equivalently by the most energetic $k^*$.\\ 
\noindent iii. $\Delta\nu_E=3,4:$ The DTOP evolution becomes progressively complex with increase of $\Delta\nu_E$. This can be seen in the cases of $\Delta\nu=3$ and $\Delta\nu=4$ for which three and four $ k^* $, and hence as many sets of critical times, exist respectively.

The above results demonstrate that DQPTs exist for the ETI system when it is quenched across phases with different winding numbers and the number of critical $ k^* $ that contribute to the dynamical free energy non-analyticities is equal to the difference of winding number of the two equilibrium phases. However, we now present another case where this one to one relation breaks down.

We carry out the first quench tabulated in Table.~1 again, with all the parameters set to the same values except $\gamma$ which we now set to a value close to zero, $\gamma=0.25$. Although the difference in $\nu_E$ is still $1$ across the phases, there now exists three $ k^* $ as can be seen clearly from Fig.~\ref{fig_except1} and Fig.~\ref{fig_except2}. It is important to note here that $\gamma=0$ is also a QCP for $b<-0.5$. The  vicinity to the QCP for $\gamma$ is the likely cause of existence of more than one $k^*$.

 Lastly, we revisit the special case outlined in Sec.~\ref{sec:ETIM}, where $\Delta\nu_E=0$ across the gap closing point, and look for DQPTs if the Hamiltonian parameter $b$ is quenched across the QCP. Remarkably, two critical time scales are observed as shown in Fig.~\ref{fig_except1} and Fig.~\ref{fig_except3}, rather than zero. Once again, we would like to point out that $\gamma=b=0$ is a multicritical point if the other parameters are held constant. Apparently, this observation also holds true for the results reported in \ct{vajna14} for DQPTs in the XY model where DQPTs were observed for quenches within the same phase.
\begin{figure*}
	\begin{subfigure}{0.4\textwidth}
		\includegraphics[width=\columnwidth]{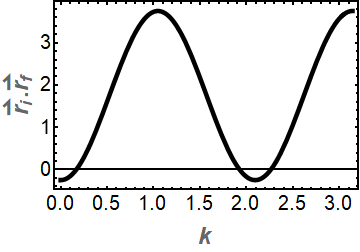}
		\caption{}
		\label{fig_k1}
	\end{subfigure}
	\begin{subfigure}{0.4\textwidth}
		\includegraphics[width=0.9\columnwidth]{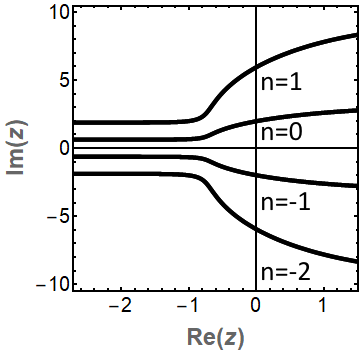}
		\caption{}
		\label{fig_f1}
	\end{subfigure}
	\begin{subfigure}{0.4\textwidth}
		\includegraphics[width=\columnwidth]{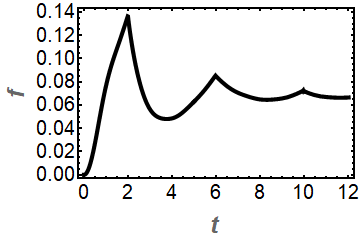}
		\caption{}
		\label{fig_e1}
	\end{subfigure}
	\begin{subfigure}{0.4\textwidth}
		\includegraphics[width=\columnwidth]{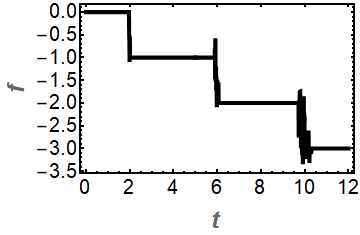}
		\caption{}
		\label{fig_d1}
	\end{subfigure}
	\cc
	\caption{ These figures represent the special situation discussed in section \ref{sec: DCM}. Although the plot $\vec{r}_i(k)\cdot\vec{r}_f(k)$ (fig (a)) shows three $k^*$, fig (b) shows that these $k^*s$ correspond to a single $t^*$. This implies that  the lines of FZs retrace their path(thrice) on the complex time plane as $k$ varies from $0$ to $\pi$. Real time evolution of free energy (c) and DTOP (d) thus show a SI like behaviour.}. 
\end{figure*}
\section{Cluster Ising Chain: degenerate critical modes }
\label{sec: DCM}
Let us recall the Hamiltonian (refer to Eq.~\eqref{eq_hamil})choosing a set of parameter values as $b=-1, \gamma=-1, \delta=1, g=0$. For this set of parameters, the ETI model reduces to the cluster Ising model described by
the Hamiltonian
\begin{equation}
H=\sum_{i=1}^N \left(a\sigma_i^y\sigma_{i+1}^y-\sigma_i^z \sigma_{i-1}^x\sigma_{i+1}^x\right)
\label{eq_cluster}
\end{equation}
The spectrum, which can be easily obtained as 
\begin{equation}
\epsilon(k)=\sqrt{1+a^2 - 2a\cos{3k}}
\label{eq_clus_spect}
\end{equation}
reveals QCPs at $a=\pm1$. Interestingly, for $a=1$, the gap in the spectrum vanishes at two modes, viz. $k=0$ and $k=2\pi/3$, simultaneously. Therefore, the system undergoes a QPT from a 'cluster' phase to an antiferromagnetic phase as $a$ is increased across the QCP. We note in passing that this model can be recast in a sum of three decoupled Ising chains \ct{smacchia11}. One can show that the change in $\nu_E$ corresponding to the QPT in the cluster Ising model is $2\rightarrow-1$, i.e. $\Delta\nu_E=3$ while in the TI model, $\Delta\nu_E=1$ across the QCP.

However, a more striking feature emerges in the DQPTs that are observed following a sudden quench in the parameter $a$ across the above mentioned QCP at $a=1$. Although there are three critical modes $k^*$ (see Fig.~\ref{fig_k1}) as expected for a $\Delta\nu_E=3$ quench, there is only one critical time scale $t^*$ in the subsequent dynamics as shown in Fig.~\ref{fig_f1}, \ref{fig_e1} and \ref{fig_d1}.  A re-look at Eq.~\eqref{eq_cr_tscale} reveals that this is only possible if the $k^*$ are degenerate. For the particular quench above, Eq.~\eqref{eq_critical_k} dictates that the $k^*$ are given by 
\begin{equation}
\cos{3k^*}=\frac{1+a_ia_f}{a_i+a_f},
\label{eq_clus_cri}
\end{equation}   
which immediately implies that the spectrum given by Eq.~\eqref{eq_clus_spect} is indeed degenerate at the critical modes $k^*$, notably for any values of $a_i$ and $a_f$ as long as they are on the two sides of the QCP. This emphasizes that this degeneracy is not accidental and is maintained irrespective of the quench amplitude as long as the RHS of Eq.~\eqref{eq_clus_cri} is less than one, which again translates to the condition that the quench is performed across the QCP; this degeneracy is an artifact of the spectrum in Eq.~\eqref{eq_clus_spect} of the cluster model that incorporated  three TI spectra embedded in it.

\section{\label{sec:con}Conclusions}

In this paper, we have investigated the role of three-spin interactions on the non-equilibrium dynamics of a transverse XY model. The situation here resembles the long-range Kitaev chain where
the long-range superconducting term leads to phases with higher winding numbers \ct{jaeger17}: Here as well, the extended nature of the model, which is manifested in  the three-spin interactions, allows for  the possibility of observing quantum phases with higher winding numbers. In the equilibrium scenario, we have demonstrated an example where the winding number does not change across a gap-closing point, although the non-equilibrium dynamics showed non-analyticities  in the form of DQPT when quenched across the same gapless  point. The DQPT in this case had contribution from two critical modes;  this illustrates the breakdown of the usual one-to-one correspondence between number of critical modes (and fundamental critical time scales) and $\Delta\nu_E$ that we established from analyzing quenches across equilibrium phases with $\Delta\nu_E=1,2,3,4$. We further demonstrated  another pathological case  where three $k^*$ were shown to exist for $\Delta\nu_E=1$. These two special cases lead us to conclude that the vicinity to multicritical points that arise due to augmented parameter space of the Hamiltonian resulting from
three spin interactions, is possibly at the root of the  breakdown of the one-to-one correspondence between $\Delta\nu_E$ and number of critical time scales. A similar observation has also been reported in \ct{dutta17, jaeger17}. Finally, we showed that in the limit where the ETI model reduces to a cluster Ising model, the critical modes become degenerate leading to a single critical time scale, unlike previously studied cases. This degeneracy, which is not accidental, has not been observed previously to the best of our knowledge and is an artifact
of the spectrum of the cluster Ising model as demonstrated  in Eqs.~\eqref{eq_clus_spect} and \eqref{eq_clus_cri}.

\bigskip

AD acknowledges SERB, DST, New Delhi for financial support. SB acknowledges CSIR, India for financial support. We also acknowledge Souvik Bandyopadhyay, Utso Bhattacharya, Sudarshana Laha and Somnath Maity for their critical comments.


\begin{thebibliography}{11}



%


\bi{greiner02} M. Greiner , O. Mandel, T.  W. Hansch and  I. Bloch,  Nature {\bf 419}, 51 (2002).    


\bi{kinoshita06} T. Kinoshita, T. Wenger and D. S. Weiss,  Nature {\bf 440}, 900 (2006).    

\bi{fausti11} D. Fausti, R. I. Tobey, , N. Dean,  S. Kaiser, A. Dienst, M. C. Hoffmann, S. Pyon, T. Takayama, H. Takagi,4, A. Cavalleri,  Science {\bf 331}, 189 (2011). 

\bi{gring12} M. Gring, M. Kuhnert, T. Langen, T. Kitagawa, B. Rauer, M. Schreitl, I. Mazets1, D. Adu Smith, E. Demler, and J. Schmiedmayer, 
Science {\bf 337}, 1318 (2012).   


\bi{trotzky12}  S. Trotzky,	Y-A. Chen,	A. Flesch,	I. P. McCulloch,	U. SchollwÂck, J. Eisert and I. Bloch, Nature {\bf 8}, 325 (2012).   





\bi{cheneau12}  M. Cheneau,	P. Barmettler,	D.  Poletti,	 M. Endres,	P.  Schauss,	T. Fukuhara,	C. Gross,	I. Bloch,	C.  Kollath	 and S.  Kuhr, Nature {\bf 481}, 484  (2012). 

\bi{rechtsman13}  M. C. Rechtsman,	J. M. Zeuner,	Y.  Plotnik,	 Y.  Lumer,	D.Podolsky,	F.  Dreisow,	S. Nolte,	M. Segev	and  A. Szameit,    Nature
{\bf 496}  196 (2013).   



\bi{schreiber15}     M.  Schreiber,  S. S. Hodgman, P.  Bordia,  Henrik P. LÂschen, M. H. Fischer, R. Vosk, E. Altman, U. Schneider, I. Bloch, Science {\bf 349}, 842 (2015).  





\bi{calabrese06} P. Calabrese, and  J. Cardy, Phys. Rev. Lett. {\bf 96}, 136801 (2006); J. Stat. Mech,
P06008 (2007).    

\bi{rigol08} M. Rigol, V. Dunjko and M. Olshanii,  Nature {\bf 452}, 854 (2008).   

\bibitem{oka09} T Oka, H Aoki, Phys.  Rev.  B {\bf 79} 081406 (2009).   

\bibitem{mukherjee09}
V. Mukherjee  and A. Dutta, J. Stat. Mech. P05005 (2009).  

\bi{bermudez09} A. Bermudez, D. Patane,  L. Amico, M. A. Martin-Delgado,  Phys. Rev. Lett. {\bf 102}, 135702, (2009).  

\bibitem{kitagawa10} T. Kitagawa, E. Berg, M. Rudner, and E. Demler,  Phys. Rev. B
{\bf 82}, 235114 (2010).  

\bibitem{das10}
A. Das,   Phys. Rev. B {\bf 82}, 172402 (2010).  

\bibitem{pal10} A Pal and DA Huse, Phys. Rev. B {\bf 82}, 174411  (2010).   

\bibitem{lindner11} N. H. Lindner, G. Refael and V. Galitski,  Nat. Phys. {\bf 7}, 490-495, (2011).  


\bibitem{thakurathi13} M. Thakurathi, A. A. Patel, D. Sen, and A. Dutta,  Phys. Rev. B {\bf 88}, 155133 (2013).  


\bibitem{Russomanno_PRL12}
{A. Russomanno,  A. Silva  and G. E. Santoro} , Phys. Rev.
Lett. {\bf 109}, 257201 (2012); S. Sharma, A. Russomanno, G. E. Santoro and A. Dutta, EPL {\bf 106},  67003 (2014).   

\bi{nag14} T. Nag, S.  Roy, A. Dutta, and D.  Sen, Phys. Rev. B {\bf 89}, 165425 (2014).

\bi{patel13} A. A. Patel, S. Sharma, A. Dutta, Eur. Phys. Jour. B {\bf 86}, 367 (2013); A. Rajak and  A. Dutta, Phys. Rev. E 89, 042125, 2014. P. D. Sacramento, Phys. Rev. E {\bf 90} 032138, (2014); M. D. Caio, N. R. Cooper and M. J. Bhaseen, Phys. Rev. Lett. {\bf 115}, 236403 (2015).    

\bi{nandkishore15} R. Nandkishore, D. A. Huse, Annual Review of Condensed Matter Physics,  {\bf 6}, 15-38 (2015).  


\bibitem{sen16} A.  Sen, S.  Nandy, K. Sengupta, Phys. Rev. B {\bf 94}, 214301 (2016).   

\bi{bukov16} 
M. Bukov, L. D'Alessio and A. Polkovnikov, Adv. Phys. {\bf 64} , No. 2, 139-226 (2016).   






\bibitem {dziarmaga10} J. Dziarmaga,  Advances in Physics  {\bf 59}, 1063 (2010).  

\bibitem{polkovnikov11} A. Polkovnikov, K. Sengupta, A. Silva, and M. Vengalattore, Rev. Mod. Phys. {\bf 83}, 863 (2011).  



\bi{dutta15} A. Dutta, G. Aeppli, B. K. Chakrabarti, U. Divakaran, T. 
Rosenbaum and D. Sen, \textit{Quantum Phase Transitions in Transverse Field 
	Spin Models: From Statistical Physics to Quantum Information} (Cambridge 
University Press, Cambridge, 2015).    

\bi{eisert15} J. Eisert, M. Friesdorf and C. Gogolin,  Nat. Phys. {\bf 11}, 124 (2015).  

\bi{alessio16} L. D'Alessio, Y.  Kafri, A. Polkovnikov, M. Rigol,   Adv. Phys. {\bf 65}, 239 (2016).  

\bi{jstat} J.  Stat.  Mech.: Theo. and Expt,   special issue{\it Quantum Integrability in Out of Equilibrium Systems} edited by  P. Calabrese., F. H. L. Essler and G. Mussardo, {\bf 2016} (2016).
 
\bi{heyl13} M. Heyl, A. Polkovnikov, and S. Kehrein,  Phys. Rev. Lett., {\bf 110}, 135704 (2013).   
  
\bibitem{sachdev96} S. Sachdev, 
(Cambridge University Press, Cambridge, UK, 2010).  

\bibitem{suzuki13} S. Suzuki, J-i Inoue and Bikas K. Chkarabarti,   (Springer, Lecture Notes in Physics, Vol. 862 (2013)).         

\bi{gambassi11} A. Gambassi and  A.  Silva, arXiv: 1106.2671 (2011); , Phys. Rev. Lett. {\bf 109}, 250602 (2012); P. Smacchia and A. Silva,  Phys. Rev. E {\bf 88}, 042109, (2013).


\bi {lee52} C. Yang and T. Lee, Phys. Rev. {\bf 87}, 404 (1952). 

\bi {fisher65} M.E. Fisher, in {\it Boulder Lectures in Theoretical Physics} (University of Colorado, Boulder, 1965), Vol. 7.     


\bi{saarloos84} W. van Saarloos and D. Kurtze, J. Phys. A {\bf 17}, 1301 (1984).   


\bi{karrasch13} C. Karrasch and D. Schuricht,  Phys. Rev. B, {\bf 87}, 195104 (2013).  

\bi{kriel14} N. Kriel, C. Karrasch, and S. Kehrein, Phys. Rev. B {\bf 90}, 125106 (2014).

\bi{andraschko14} F. Andraschko, J. Sirker, Phys. Rev. B {\bf 89}, 125120 (2014).  

\bi{canovi14} E. Canovi, P. Werner, and M. Eckstein, Phys. Rev. Lett. {\bf 113}, 265702 (2014).

\bi{heyl14} M. Heyl, Phys. Rev. Lett., {\bf 113}, 205701 (2014).

\bi{heyl15} M. Heyl, Phys. Rev. Lett., {\bf 115}, 140602 (2015) .

\bi{budich15} J. C. Budich and  M. Heyl,  Phys. Rev. B {\bf 93}, 085416 (2016).     



\bi{palami15} T. Palmai, Phys. Rev. B {\bf 92}, 235433 (2015).  

\bi{divakaran16} U. Divakaran, S. Sharma and A. Dutta, Phys. Rev. E {\bf 93}, 052133 (2016). 


\bi{huang16} Z.  Huang, and A.  V. Balatsky, Phys. Rev. Lett. {\bf 117}, 086802 (2016).   

\bi{puskarov16} T. Puskarov and D. Schuricht,  arXiv: 1608.05584 (2016).  

\bi{zhang16} J. M. Zhang abd  H.-T. Yang, arXiv: 1605.05403 (2016). 

\bi{heyl16} M. Heyl, Phys. Rev. B {\bf 95}, 060504 (2017). 


\bi{zunkovic16} Bojan Zunkovic, Markus Heyl, Michael Knap, Alessandro Silva, arXiv:1609.08482 (2016). 


\bi{sei17} T.  Obuchi, S. Suzuki, K. Takahashi, arXiv:1702.05396 (2017). 

\bi{fogarty17} Thoms Fogarty, Ayaka Usui, Thomas Busch, Alessandro Silva, John Goold,  New J. Phys. 19 113018 (2017). 

\bi{heyl18} Daniele Trapin, Markus Heyl, arXiv:1802.00020 (2018).

\bi{heyl17} Markus Heyl,  arXiv:1709.07461 (2017)

\bi{victor17} Victor Gurarie, 10.1103/Physics.10.95 (2017)

\bi{zvyagin17} A.A. Zvyagin, Low Temp. Phys. {\bf 42}, 971 (2016).

\bi{vajna14} S. Vajna and B. Dora,  Phys. Rev. B {\bf 89}, 161105(R) (2014). 

\bi{sharma15} S. Sharma, S. Suzuki and A. Dutta,  Phys. Rev. B {\bf 92}, 104306 (2015). 


\bi{pollmann10} F. Pollmann, S. Mukerjee, A. G. Green, and J. E. Moore, Phys. Rev. E {\bf 81}, 020101(R) (2010).  

\bi{sharma16} S. Sharma, U. Divakaran, A. Polkovnikov and A. Dutta, Phys. Rev. B {\bf 93}, 144306 (2016).  



\bi{vajna15} S. Vajna and B. Dora, Phys. Rev. B {\bf 91}, 155127 (2015). 



\bi{schmitt15} M. Schmitt and S. Kehrein, Phys. Rev. B {\bf 92}, 075114 (2015). 

\bi{bhattacharya1} Utso Bhattacharya and Amit Dutta, Phys. Rev. B {\bf96}, 014302 (2017). 


\bi{bhattacharya2} Utso Bhattacharya and Amit Dutta, Phys. Rev. B 95, 184307 (2017).

\bi{utso} U. Bhattacharya, S. Bandyopadhyay and A. Dutta, Phys. Rev. B {\bf96}, 180303(R) (2017).    

\bi{heyl} M.Heyl and J.C. Budich, Phys. Rev. B {\bf96}, 180304(R) (2017) . 



\bi{flaschner} N. Flaschner, D. Vogel, M. Tarnowski, B, S. Rem, D.-S. Luhmann, M. Heyl, J. Budich, L. Mathey, K. Sengstock,
C. Weitenberg, Nat. Phys. 1745-2481 (2017). 


\bi{jurcevic16} P. Jurcevic, H. Shen, P. Hauke, C. Maier, T. Brydges, C. Hempel, B. P. Lanyon, M. Heyl, R. Blatt, C. F. Roos, Phys. Rev. Lett. {\bf119}, 080501 (2017). 



\bi{zhang} G. Zhang and Z. Song, Phys. Rev. Lett. {\bf115}, 177204 (2015). 

\bi{zhang17} XZ. Zhang, JL. Guo, Quantum Inf Process, {\bf16:223} (2017).


\bi{titvinidze03} I. Titvinidze and G. I. Japaridze, Eur. Phys. J. B {\bf32}, 383 (2003); A. A. Zvyagin and G. A. Skorobagat’ko, Phys. Rev. B {\bf73}, 024427 (2006).

\bi{fabrizio96} M. Fabrizio, Phys.Rev.B 54, 10054 (1996); R. Arita, K. Kuroki, H. Aoki, and M. Fabrizio, \textit{ibid.}57, 10324 (1998).

\bi{suzuki71} M. Suzuki, Phys. Lett. A 34, 338 (1971); Prog. Theor. Phys.46, 1337 (1971).

\bi{perk75} J.H.H.Perk, H.W.Capel, M.J.Zuilhof and Th.J.Siskens, Physica A 81, 319 (1975).

\bi{divakaran13} U. Divakaran, Phys.Rev.E {\bf 88}, 052122 (2013).


\bi{chowdhury10} D. Chowdhury, U. Divakaran, and A. Dutta, Phys. Rev. E {\bf 81}, 012101 (2010).

\bi{vodola14} Davide Vodola, Luca Lepori, Elisa Ercolessi, Alexey V. Gorshkov, and Guido Pupillo
Phys. Rev. Lett. {\bf 113}, 156402 (2014).

\bi{vodola16} D. Vodola, L. Lepori, E. Ercolessi, and G. Pupillo, New J.Phys. {\bf 18}, 015001 (2016).

\bi{viyuela16} O. Viyuela, D. Vodola, G. Pupillo, and M. A. Martin-Delgado, Phys. Rev. B {\bf 94}, 125121 (2016).

\bi{regemortel16} M. V. Regemortel, M. Wouters, and D. Sels,Phys. Rev. A {\bf 93}, 032311 (2016).

\bi{lepori17} L. Lepori, A. Trombettoni, and D. Vodola, J. Stat. Mech. 033102 (2017).

\bi{dutta17} Anirban Dutta and Amit Dutta, Phys. Rev. B {\bf 96}, 125113 (2017).

\bi{jaeger17} P. Jaeger, S. Eggert, J. Sirker, Diplomarbeit, Bulk-boundary correspondence in non-equilibrium dynamics of one-dimensional topological insulators, University of Manitoba (2017).

\bi{sedlmayr17} N. Sedlmayr, P. Jäger, M. Maiti, J. Sirker, arXiv:1712.03618 (2017).

\bi{dutta01} A. Dutta, J.K. Bhattacharjee, Phys. Rev. B {\bf64}, 184106 (2001).

\bi{halimeh17} J. C. Halimeh and V. Zauner-Stauber, Phys. Rev. B {\bf96}, 134427 (2017).

\bi{homri17} I. Homrighausen, N. O. Abeling, V. Zauner-Stauber, and J. C.Halimeh, Phys.Rev.B {\bf 96}, 104436 (2017).

\bi{zunkovicb} B. Zunkovic, A. Silva, M. Fabrizio, Philos Trans A Math Phys Eng Sci. 374(2069), (2016).

\bi{son11} W.  Son, L. Amico, R. Fazio, A. Hamma, S.  Pascazio, Vl. Vedral, Europhys. Lett. vol. {\bf 95}, 50001 (2011)

\bi{smacchia11} P.  Smacchia, L. Amico, P. Facchi, R. Fazio, G. Florio, S. Pascazio, V.  Vedral, 
Phys.Rev.A {\bf 84} 022304, (2011). 

































































\end{thebibliography}
\end{document}